\documentclass[aps,showpacs,superscriptaddress,twocolumn]{revtex4}%
\usepackage{amsfonts}
\usepackage{amsmath}
\usepackage{amssymb}
\usepackage{graphicx}%
\providecommand{\U}[1]{\protect\rule{.1in}{.1in}}

\begin{document}
\title{Topological Superfluid Transition Induced by Periodically Driven Optical Lattice}
\author{Guocai Liu}
\affiliation{School of Science, Hebei University of Science and Technology, Shijiazhuang
050018, China}
\author{Ningning Hao}
\affiliation{Beijing National Laboratory for Condensed Matter Physics, Institute of
Physics, Chinese Academy of Science, Beijing 100190, China}
\affiliation{Department of Physics, Purdue University, West Lafayette, Indiana 47907, USA}
\author{Shi-Liang Zhu}
\affiliation{Laboratory of Quantum Information Technology, School of Physics and
Telecommunication Engineering, South China Normal University, Guangzhou, China}
\affiliation{Center for Quantum Information, IIIS, Tsinghua University, Beijing, China}
\author{W. M. Liu}
\affiliation{Beijing National Laboratory for Condensed Matter Physics, Institute of
Physics, Chinese Academy of Science, Beijing 100190, China}


\begin{abstract}
We propose a scenario to create topological superfluid in a periodically
driven two-dimensional square optical lattice. We study the phase diagram of a
spin-orbit coupled s-wave pairing superfluid in a periodically driven
two-dimensional square optical lattice. We find that a phase transition from a
trivial superfluid to a topological superfluid occurs when the potentials of
the optical lattices are periodically changed. The topological phase is called
Floquet topological superfluid and can host Majorana fermions.

\end{abstract}
\maketitle

\section{INTRODUCTION}

Optical lattice system has gradually become a promising platform to study
many-body quantum systems because of lots of significant advances in cold-atom
experiments \cite{Lewe}. In particular, recent theoretical and experimental
progress in laser-induced-gauge-field \cite{Ruse,Oste,Juze,Zhu2,Lin1,Lin2}
makes it a hot spot to study topological quantum states in cold atoms system
\cite{Stanescu1,Bermudez,Satija,LXJ2}. Recently, topological quantum states
have attracted considerable interest in condensed physics; however, subject to
the compounds' natural properties \cite{Stan}, we have to rely on serendipity
in looking for topological materials in solid-state structures
\cite{Mura,Koni,Hsieh3,Zhang}. In contrast, one can engineer the Hamiltonian
of an optical lattice system to realize variant quantum phase states
\cite{Gold2,Zhu3,Liu}.

In this paper, we show that periodically driven perturbations may give rise to
a phase transition from a trivial superfluid to a topological one, which
carries the hallmark with topological protected gapless edges on the
boundaries of the system. Time-periodic dependent Hamiltonian can be described
by Floquet's theorem, which is used to explain quantized adiabatic pumping
phenomena \cite{Fu,Inoue1,Thouless,Niu}. Recently, it demonstrated that the
phase transition from a superfluid to a Mott insulator in one-dimensional
Bose-Hubbard model can be induced by a periodically driven optical lattice
\cite{Ecka}. We extend this phase transition mechanism to explore the
topological phase transition in a two-dimensional optical lattice. We study
the phase diagram of a spin-orbit coupled s-wave pairing superfluid in a
periodically driven two-dimensional (2D) square optical lattice. We find that
a topological phase transition from a trivial superfluid to a topological
superfluid can be induced in periodically modulated optical lattices. The
topological phase is called Floquet topological superfluid
\cite{Jiang,Kita,Inoue2,Lind,Kita2} and can host Floquet Majorana fermions. It
was proposed that a topological phase can be realized in a BCS s-wave
superfluid of ultracold fermionic atoms in the presence of both a Rashba
spin-orbit (SO) interaction and a large perpendicular Zeeman field
\cite{Sato,ZhangPRL,ZhangPRA,Zhu4}; however, the Rashba spin-orbit coupling
and a large perpendicular Zeeman field are hard to be simultaneously realized
for cold fermionic atoms \cite{ZhangPRA,Zhu4}. We will prove that if one
replaces the Zeeman field by a periodically driven optical lattice
\cite{Ecka}, a spin-orbit coupled BCS s-wave superfluid will still allow a
realization of topological superfluid through modifying the oscillating
amplitude (or modulation strength) of optical lattice. Therefore, we provide
an alternative method to create an important topological superfluid which can
host Majorana fermions.

The paper is organized as follows: In Sec. II, we introduce the s-wave
superfluid model in a square optical lattice in the presence of both a Rashba
SO coupling and a periodically modulated optical lattice potential. A
Zeeman-magnetic-field-like term will be derived under the first order
approximation; In Sec. III we present a two-band approximation and explain the
topological phase transition at the $\Gamma$ point in the first Brillouin zone
(BZ). At last, we give a brief summary in Sec. IV.

\section{MODEL}

The tight-binding Hamiltonian, which describes an s-wave superfluid of neutral
fermionic atoms in a 2D optical square lattice, is given by
\begin{equation}
H\left(  t\right)  =H_{0}+H_{d}\left(  t\right)  , \label{E1}%
\end{equation}
where%
\begin{align}
H_{0}  &  =-t\sum_{\langle ij\rangle}c_{i}^{\dag}c_{j}-i\lambda\sum_{\langle
ij\rangle}c_{i}^{\dag}\left(  \mathbf{\sigma\times\hat{d}}_{ij}\right)
_{z}c_{j}\nonumber\\
&  +\mu\sum_{i}c_{i}^{\dag}c_{i}+U\sum_{i}c_{i\uparrow}^{\dag}c_{i\downarrow
}^{\dag}c_{i\downarrow}c_{i\uparrow}, \label{E2}%
\end{align}
and%
\begin{equation}
H_{d}\left(  t\right)  =\mathbf{K}\left(  t\right)  \cdot\sum_{i}%
\mathbf{r}_{i}c_{i}^{\dag}c_{i}. \label{E3}%
\end{equation}
Here $t$ is the hopping amplitude between the nearest neighbor link $\langle
i,j\rangle$, $c_{i}^{\dag}$=$\left(  c_{i\uparrow}^{\dag},c_{i\downarrow
}^{\dag}\right)  $ with $c_{i\alpha}^{\dag}$ ($c_{i\alpha}$) denoting the
creation (annihilation) operator of a fermionic atom with pseudospin $\alpha$
(up or down) on lattice site $i$. The second term in Eq. (\ref{E2}) represents
a Rashba SO coupling interaction which can be obtain by
laser-induced-gauge-field method, $\lambda$ is the coupling
coefficient,\textbf{ }$\mathbf{\sigma}$ are the Pauli matrices and
$\mathbf{\hat{d}}_{ij}$ is a unit vector along the bond that connects site $j$
to $i$. $\mu$ is the chemical potential and $U<0$ denotes an on-site
attractive interaction which is easy to obtain via an s-wave Feshbach
resonance in cold atom system. The oscillating Hamiltonian $H_{d}$, with
$\mathbf{K}\left(  t\right)  =K\left(  \cos\left(  \omega t\right)
,\sin\left(  \omega t\right)  \right)  $, mimics a monochromatic electric
dipole potential with frequency $\omega$ and amplitude $K$. This term can be
realized experimentally by periodically shifting the position of a mirror
employed to generate the standing laser waves along $x$- and $y$-directions,
and transforming to the comoving frame of reference \cite{Ecka}. We choose
$t$=$1$ as the energy unit and the distance $a$ between the nearest sites as
the length unit throughout this paper. It was demonstrated in Ref. \cite{Sato}
that the Hamiltonian $H_{0}$ in a mean field approximation and combination
with a perpendicular Zeeman field can support a topological superfluid. On the
other hand, replaced the Hamiltonian $H_{0}$ with an one-dimensional
Bose-Hubbard Hamiltonian, it was shown in Ref. \cite{Ecka} that a phase
transition from a superfluid to a Mott insulator can be induced by $H_{d}$ in
its one-dimensional form.

When a Hamiltonian of quantum system has a periodic dependence on time, i.e.,
$H\left(  t\right)  $=$H\left(  t+T\right)  $ with period $T$=$2\pi/\omega$,
the Hamiltonian satisfies the discrete time translational symmetry,
$t\rightarrow t+T$, which can been described by Floquet's theorem
\cite{Ecka,Shirl,Grif}. Floquet's theorem tell us that the Schr\"{o}dinger
equation with time-periodic dependent Hamiltonian has a complete set of
solutions with the form $\left\vert \psi_{n}\left(  t\right)  \right\rangle
$=$\left\vert u_{n}\left(  t\right)  \right\rangle \exp\left(  -i\varepsilon
_{n}t/\hbar\right)  $. Here, the periodic function $\left\vert u_{n}\left(
t\right)  \right\rangle =\left\vert u_{n}\left(  t+T\right)  \right\rangle $,
an analog of Bloch states known from spatially periodic crystals, satisfies
the eigenvalue equation%
\begin{equation}
\left[  H\left(  t\right)  -i\hbar\partial_{t}\right]  \left\vert u_{n}\left(
t\right)  \right\rangle =\varepsilon_{n}\left\vert u_{n}\left(  t\right)
\right\rangle . \label{E4}%
\end{equation}
We call $\mathcal{H}\left(  t\right)  $=$H\left(  t\right)  -i\hbar
\partial_{t}$ as the Floquet Hamiltonian and the eigenvalues $\varepsilon_{n}$
as quasienergies which are defined modulo the frequency $\omega$=$2\pi/T$.

The Floquet basis%
\begin{equation}
\left\vert \left\{  n_{i}\right\}  ,m\right\rangle =\left\vert \left\{
n_{i}\right\}  \right\rangle \exp\left[  \mathtt{-}\frac{i}{\hbar\omega}%
\int_{-\infty}^{t}dt^{\prime}\mathbf{K}\left(  t^{\prime}\right)
\mathtt{\cdot}\sum_{i}\mathbf{r}_{i}n_{\mathbf{i}}\mathtt{+}im\omega t\right]
, \label{E5}%
\end{equation}
where $\left\vert \left\{  n_{i}\right\}  \right\rangle $ indicates a Fock
state with $n_{i}$ particles on the $i$th site, and $m$ accounts for the zone
structure \cite{Ecka}, consist of an extended Hilbert space of $T$-periodic
functions with the scalar product given by%
\begin{equation}
\left\langle \left\langle \cdot\mathtt{\mid}\cdot\right\rangle \right\rangle
=\frac{1}{T}\int_{0}^{T}dt\left\langle \cdot\mathtt{\mid}\cdot\right\rangle ,
\label{E6}%
\end{equation}
i.e., by the usual scalar product $\left\langle \cdot\mathtt{\mid}%
\cdot\right\rangle $ combined with time-averaging. Hence, the quasienergies
are obtained by computing the matrix elements of the Floquet operator
$H\left(  t\right)  -i\hbar\partial_{t}$ in the basis (\ref{E5}) with respect
to the scalar product (\ref{E6}), and diagonalizing. By a straightforward
calculation, we can obtain the matrix elements of some operators in Floquet
Hamiltonian $\mathcal{H}\left(  t\right)  $:%
\begin{equation}
\left\langle \left\langle \left\{  n_{i}^{\prime}\right\}  ,m^{\prime
}\right\vert c_{i\alpha}^{\dag}c_{j\alpha^{\prime}}\left\vert \left\{
n_{i}\right\}  ,m\right\rangle \right\rangle =e^{-i\left(  m^{\prime
}-m\right)  \theta_{ij}}J_{m^{\prime}-m}\left(  z_{ij}\right)  , \label{E7}%
\end{equation}%
\begin{equation}
\left\langle \left\langle \left\{  n_{i}^{\prime}\right\}  ,m^{\prime
}\right\vert c_{i\alpha}^{\dag}c_{i\alpha}\left\vert \left\{  n_{i}\right\}
,m\right\rangle \right\rangle =n_{i\alpha}\delta_{m,m^{\prime}}, \label{E8}%
\end{equation}%
\begin{equation}
\left\langle \left\langle \left\{  n_{i}^{\prime}\right\}  ,m^{\prime
}\right\vert c_{i\uparrow}^{\dag}c_{i\downarrow}^{\dag}c_{i\downarrow
}c_{i\uparrow}\left\vert \left\{  n_{i}\right\}  ,m\right\rangle \right\rangle
=n_{i\uparrow}n_{i\downarrow}\delta_{m,m^{\prime}}, \label{E9}%
\end{equation}
where $J_{m^{\prime}-m}\left(  z_{ij}\right)  $\ is the Bessel function of the
$\left(  m^{\prime}-m\right)  $th order and $z_{ij}$=$\frac{K}{\hbar\omega
}\sqrt{x_{ij}^{2}+y_{ij}^{2}}$. Here, $x_{ij}=\left(  \mathbf{r}_{i}\right)
_{x}-\left(  \mathbf{r}_{j}\right)  _{x}$, $y_{ij}=\left(  \mathbf{r}%
_{i}\right)  _{y}-\left(  \mathbf{r}_{j}\right)  _{y}$ and $\tan\theta
_{ij}=x_{ij}/y_{ij}$. In the above matrix, the diagonal block of the Floquet
Hamiltonian, $\mathcal{H}^{\left(  mm\right)  }$, is the $n$-photon sector,
i.e., the subspace with $n$ photons and the non-diagonal blocks $\mathcal{H}%
^{\left(  m^{\prime}m\right)  }$ with $m^{\prime}\neq m$ correspond to the
interaction between different subspaces \cite{Inoue2}. For sufficiently high
frequencies, we can argue, from Eq. (\ref{E7}-\ref{E9}), that the driven
system (\ref{E1}) behaves similar to the undriven system (\ref{E2}), but with
the tunneling matrix element $t$ and the SO coupling $\lambda$ of the latter
being replaced by the effective matrix element $t\sim tJ_{0}\left(
z_{ij}\right)  $ and $\lambda\sim\lambda J_{0}\left(  z_{ij}\right)  $,
respectively. Now, suppose that we enhance the modulation strength $K$, then
we have to consider the coupling of other photon sectors. For simplify, we
only consider coefficient of subspace with $n=1$ photon on the subspace with
$n=0$ photon. When $K$ is strong enough but still satisfy $z_{ij}<<1$, the
system has the effective Hamiltonian \cite{Kita2}%
\begin{equation}
\mathcal{H}_{eff}=\mathcal{H}^{\left(  00\right)  }+\frac{1}{\hbar\omega
}\left[  \mathcal{H}^{-1},\mathcal{H}^{+1}\right]  . \label{E10}%
\end{equation}
Here, $\mathcal{H}^{-1}$ ($\mathcal{H}^{+1}$) denotes the non-diagonal block
with $m\prime-m=-1$ $\left(  +1\right)  $ around the $0$-photon sector.
According to Eqs. (\ref{E7}-\ref{E9}), we can obtain%

\begin{align}
\mathcal{H}^{\left(  00\right)  }  &  =-t\sum_{\langle ij\rangle}J_{0}\left(
z_{ij}\right)  c_{i}^{\dag}c_{j}-i\lambda\sum_{\langle ij\rangle}J_{0}\left(
z_{ij}\right)  c_{i}^{\dag}\left(  \mathbf{\sigma\times\hat{d}}_{ij}\right)
_{z}c_{j}\nonumber\\
&  +\mu\sum_{i}c_{i}^{\dag}c_{i}+\psi_{s}\sum_{i}\left(  c_{i\uparrow}^{\dag
}c_{i\downarrow}^{\dag}+\text{H.c.}\right)  , \label{E11}%
\end{align}%
\begin{align}
\mathcal{H}^{-1}  &  =-t\sum_{\langle ij\rangle}e^{i\theta_{ij}}J_{-1}\left(
z_{ij}\right)  c_{i}^{\dag}c_{j}\nonumber\\
&  -i\lambda\sum_{\langle ij\rangle}e^{i\theta_{ij}}J_{-1}\left(
z_{ij}\right)  c_{i}^{\dag}\left(  \mathbf{\sigma\times\hat{d}}_{ij}\right)
_{z}c_{j}, \label{E12}%
\end{align}%
\begin{align}
\mathcal{H}^{+1}  &  =-t\sum_{\langle ij\rangle}e^{-i\theta_{ij}}J_{+1}\left(
z_{ij}\right)  c_{i}^{\dag}c_{j}\nonumber\\
&  -i\lambda\sum_{\langle ij\rangle}e^{-i\theta_{ij}}J_{+1}\left(
z_{ij}\right)  c_{i}^{\dag}\left(  \mathbf{\sigma\times\hat{d}}_{ij}\right)
_{z}c_{j}. \label{E13}%
\end{align}
In the derivation of Eq. (\ref{E11}), we have made a mean field approximation
and $\psi_{s}$ is the gap function.

By using the Fourier transform of atomic operators $c_{i\sigma}^{\dag}$, i.e.,%
\begin{equation}
\ c_{i\sigma}^{\dag}=\frac{1}{\sqrt{N}}\sum_{\mathbf{k}}c_{\mathbf{k}\sigma
}^{\dag}e^{-i\mathbf{k\cdot R}_{i}}, \label{E14}%
\end{equation}
the Hamiltonian (\ref{E10}) in square lattice system can be rewritten in the
momentum space as%
\begin{equation}
\mathcal{H}_{eff}=\sum_{\mathbf{k}}\psi_{\mathbf{k}}^{+}(\mathcal{H}%
_{eff}(\mathbf{k})\psi_{\mathbf{k}}, \label{E15}%
\end{equation}
where we have defined the four-component basis operator $\psi_{\mathbf{k}}%
$=$(c_{\mathbf{k}\uparrow},c_{\mathbf{k}\downarrow},c_{-\mathbf{k}\uparrow
}^{\dag},c_{-\mathbf{k}\downarrow}^{\dag})^{\text{T}}$. The effective
Hamiltonian in momentum space is given by

\begin{widetext}
\begin{equation}
\mathcal{H}_{eff}(\mathbf{k})=\left(
\begin{array}
[c]{cccc}%
\varepsilon_{\mathbf{k}}-\Gamma\left(  \mathbf{k},z\right)  & 2\lambda
J_{0}\left(  z\right)  \alpha\left(  \mathbf{k}\right)  & 0 & \psi_{s}\\
2\lambda J_{0}\left(  z\right)  \alpha^{\ast}\left(  \mathbf{k}\right)  &
\varepsilon_{\mathbf{k}}+\Gamma\left(  \mathbf{k},z\right)  & -\psi_{s} & 0\\
0 & -\psi_{s} & -\varepsilon_{-\mathbf{k}}+\Gamma\left(  \mathbf{k},z\right)
& 2\lambda J_{0}\left(  z\right)  \alpha^{\ast}\left(  \mathbf{k}\right) \\
\psi_{s} & 0 & 2\lambda J_{0}\left(  z\right)  \alpha\left(  \mathbf{k}\right)
& -\varepsilon_{-\mathbf{k}}-\Gamma\left(  \mathbf{k},z\right)
\end{array}
\right)  , \label{E16}%
\end{equation}
\end{widetext}

where $\alpha\left(  \mathbf{k}\right)  =\sin k_{y}+i\sin k_{x},z=\frac
{Ka}{\hbar\omega},\Gamma\left(  \mathbf{k},z\right)  =\frac{16\lambda
^{2}J_{+1}\left(  z\right)  J_{-1}\left(  z\right)  }{\hbar\omega}\cos
k_{x}\cos k_{y}$, and $\varepsilon_{\mathbf{k}}=-2tJ_{0}\left(  z\right)
\left(  \cos k_{x}+\cos k_{y}\right)  -\mu$. Following the method outlined in
Ref. \cite{Sato}, one can obtain a \textquotedblleft dual\textquotedblright\ Hamiltonian

\begin{widetext}
\begin{equation}
\mathcal{H}^{D}\left(  \mathbf{k}\right)  =\left(
\begin{array}
[c]{cccc}%
\psi_{s}-\Gamma\left(  \mathbf{k},z\right)  & 2\lambda J_{0}\left(  z\right)
\alpha\left(  \mathbf{k}\right)  & 0 & -\varepsilon_{\mathbf{k}}\\
2\lambda J_{0}\left(  z\right)  \alpha^{\ast}\left(  \mathbf{k}\right)  &
-\psi_{s}+\Gamma\left(  \mathbf{k},z\right)  & \varepsilon_{\mathbf{k}} & 0\\
0 & \varepsilon_{\mathbf{k}} & \psi_{s}+\Gamma\left(  \mathbf{k},z\right)  &
-2\lambda J_{0}\left(  z\right)  \alpha^{\ast}\left(  \mathbf{k}\right) \\
-\varepsilon_{\mathbf{k}} & 0 & -2\lambda J_{0}\left(  z\right)  \alpha\left(
\mathbf{k}\right)  & -\psi_{s}-\Gamma\left(  \mathbf{k},z\right)
\end{array}
\right)  , \label{E17}%
\end{equation}
\end{widetext}

where the unitary transformation $\mathcal{H}^{D}\left(  \mathbf{k}\right)
=DH_{eff}D^{\dag}$ with%
\[
D=\frac{1}{\sqrt{2}}\left(
\begin{array}
[c]{cccc}%
1 & 0 & 0 & 1\\
0 & 1 & -1 & 0\\
0 & 1 & 1 & 0\\
-1 & 0 & 0 & 1
\end{array}
\right)  .
\]
It is easy to obtain the eigenvalues of Eq. (\ref{E17}) with%
\begin{align*}
E_{1}  &  =-\sqrt{f_{1}+f}_{2};E_{2}=-\sqrt{f_{1}-f_{2}},\\
E_{3}  &  =+\sqrt{f_{1}-f_{2}};E_{4}=+\sqrt{f_{1}+f_{2}},
\end{align*}
where we have defined%
\begin{align*}
f_{1}  &  =\Gamma^{2}\left(  \mathbf{k},z\right)  +\left(  \varepsilon
_{\mathbf{k}}^{2}+\psi_{s}^{2}\right)  +4J_{0}^{2}\left(  z\right)
\lambda^{2}\left\vert \alpha\left(  \mathbf{k}\right)  \right\vert ^{2},\\
f_{2}  &  =2\sqrt{\Gamma^{2}\left(  \mathbf{k},z\right)  \left(
\varepsilon_{\mathbf{k}}^{2}+\psi_{s}^{2}\right)  +4\lambda^{2}J_{0}%
^{2}\left(  z\right)  \varepsilon_{\mathbf{k}}^{2}\left\vert \alpha\left(
\mathbf{k}\right)  \right\vert ^{2}}.
\end{align*}
It is obvious that if only $\psi_{s}\neq0$, i.e., the system lies in the
superfluid phase, the energy levels $E_{1}$ and $E_{4}$, denoting the lowest
and the highest band, will not touch each other. Next, we discuss the levels
$E_{1}$ and $E_{2}$ (or $E_{3}$ and $E_{4}$). These two levels can touch each
other if only the following two relations
\[
\Gamma^{2}\left(  \mathbf{k},z\right)  =0
\]
and%
\[
J_{0}^{2}\left(  z\right)  \varepsilon_{\mathbf{k}}^{2}\left(  \sin^{2}%
k_{x}+\sin^{2}k_{y}\right)  =0.
\]
are simultaneously satisfied. From $\Gamma^{2}\left(  \mathbf{k},z\right)
=0$, we have $k_{x}=\pm\frac{\pi}{2}$ or $k_{y}=\pm\frac{\pi}{2}$ in the first
BZ. So, when $J_{0}\left(  z\right)  \neq0$, we have%
\[
-\mu=2tJ_{0}\left(  z\right)  \left(  \cos k_{x}+\cos k_{y}\right)  .
\]
When $-\mu>\left\vert 2tJ_{0}\left(  z\right)  \right\vert $, the levels
$E_{1}$ and $E_{2}$ (or $E_{3}$ and $E_{4}$) will not touch each other for
ever. Else, when $-\left\vert 2tJ_{0}\left(  z\right)  \right\vert \leq
-\mu\leq\left\vert 2tJ_{0}\left(  z\right)  \right\vert $, $E_{1}$ and $E_{2}$
(or $E_{3}$ and $E_{4}$) will touch each other at points $\left(  k_{x}%
=\pm\frac{\pi}{2},k_{y}=\arccos\left(  \frac{-\mu}{2tJ_{0}\left(  z\right)
}\right)  \right)  $ and $\left(  k_{x}=\arccos\left(  \frac{-\mu}%
{2tJ_{0}\left(  z\right)  }\right)  ,k_{y}=\pm\frac{\pi}{2}\right)  $. In the
following, we only consider the case, $-\mu>\left\vert 2tJ_{0}\left(
z\right)  \right\vert $, which means bands $E_{1}$ and $E_{4}$ will be off
away from the other two levels $E_{2}$ and $E_{3}$, and then the topological
properties of bands $E_{1}$ and $E_{4}$ will not change if we vary some
parameters. Therefore, we will only consider the topological properties of
band $E_{2}$ and $E_{3}$, which have chances to contact each other at the high
symmetry points, $\mathbf{K}_{i=1,...,4}=\left(  0,0\right)  ;\left(
0,\pi\right)  ;\left(  \pi,0\right)  ;\left(  \pi,\pi\right)  $ in the first
BZ, when satisfying the condition%
\begin{equation}
\Gamma^{2}\left(  z\right)  =\varepsilon_{\mathbf{K}_{i}}^{2}+\psi_{s}^{2},
\label{E18}%
\end{equation}
because band-gap closing is an essential condition for the topological
characteristic changes. We denote $\Gamma_{\mathbf{K}_{i}}^{2}\left(
\mathbf{K}_{i},z\right)  =\left(  16\lambda^{2}J_{+1}^{2}\left(  z\right)
/\hbar\omega\right)  ^{2}=\Gamma^{2}\left(  z\right)  $. Considering $z<<1$
and $-\mu>\left\vert 2tJ_{0}\left(  z\right)  \right\vert $, the two bands can
only touch at point $\mathbf{K}_{1}=\left(  0,0\right)  $ when varying the
parameter $z$.

\section{Topological phase transition}

We now study the topological properties of these two bands by two-band
approximation at point $\mathbf{K}_{1}=\left(  0,0\right)  $ in the first BZ.
We will not consider the other points since the gaps at other high symmetry
points will not shut down, leaving no influence to the topological changes. To
have a basic idea of the topological features of the system, we explore it by
using a two-band approximation at point $\mathbf{K}_{1}$ in the first BZ. We
expand the Hamiltonian (\ref{E17}) at point $\mathbf{K}_{1}$ and obtain

\begin{widetext}
\begin{equation}
\mathcal{H}_{\mathbf{K}_{1}}^{D}\left(  \mathbf{q}\right)  =\left(
\begin{array}
[c]{cccc}%
\psi_{s}+\Gamma\left(  z\right)  & 2\lambda J_{0}\left(  z\right)  q_{+} & 0 &
4tJ_{0}\left(  z\right)  +\mu\\
2\lambda J_{0}\left(  z\right)  q_{-} & -\left(  \psi_{s}+\Gamma\left(
z\right)  \right)  & -\left(  4tJ_{0}\left(  z\right)  +\mu\right)  & 0\\
0 & -\left(  4tJ_{0}\left(  z\right)  +\mu\right)  & \psi_{s}-\Gamma\left(
z\right)  & -2\lambda J_{0}\left(  z\right)  q_{-}\\
4tJ_{0}\left(  z\right)  +\mu & 0 & -2\lambda J_{0}\left(  z\right)  q_{+} &
-\left(  \psi_{s}-\Gamma\left(  z\right)  \right)
\end{array}
\right)
\end{equation}
\end{widetext}

with $q_{\pm}$=$q_{y}\pm iq_{x}$. Because $\psi_{s}<0$ and $\Gamma\left(
z\right)  >0$, when taking $\left\vert 4tJ_{0}\left(  z\right)  +\mu
\right\vert \ll1$, we can see that the major contribution to bands $E_{2}$ and
$E_{3}$ comes from the up-diagonal sector in above matrix and thus we may
treat the others as a perturbation. Under this condition, we obtain an
effective two-band Hamiltonian given by%
\begin{equation}
\mathcal{H}_{eff}^{\mathbf{K}_{1}}\left(  \mathbf{q}\right)  =2\lambda
J_{0}\left(  z\right)  q_{y}\sigma_{x}-2\lambda J_{0}\left(  z\right)
q_{x}\sigma_{y}+M\left(  z\right)  \sigma_{z}, \label{E19}%
\end{equation}
where the corresponding mass term $M\left(  z\right)  $=$\psi_{s}%
+\Gamma\left(  z\right)  +\left(  4tJ_{0}\left(  z\right)  +\mu\right)
^{2}/\left(  \psi_{s}-\Gamma\left(  z\right)  \right)  $ and $\sigma
_{\nu=x,y,z}$ the Pauli matrices. Let $M\left(  z\right)  $=$0$, we obtain the
gapless condition Eq. (\ref{E18}) again at $\mathbf{K}_{1}$ point.
Eq.(\ref{E19}) can be written as $\mathcal{H}_{eff}^{\mathbf{K}_{1}}\left(
\mathbf{q}\right)  ={\mathbf{\sigma}}\cdot\mathbf{d}$, where the vector
$\mathbf{d}=\{2\lambda J_{0}\left(  z\right)  q_{y},-2\lambda J_{0}\left(
z\right)  q_{x},M\left(  z\right)  \}$. The topological features of the system
can be characterized by the winding number ( first Chern number) of the Berry
phase gauge field $C=\frac{1}{4\pi}\int dk_{x}\int dk_{y}\mathbf{\hat{d}}%
\cdot(\frac{\partial\mathbf{\hat{d}}}{\partial{k_{x}}}\times\frac
{\partial\mathbf{\hat{d}}}{\partial k_{y}})$ in the first Brillouin zone,
where $\mathbf{\hat{d}}=\mathbf{d}/|\mathbf{d}|$. When $M\left(  z\right)
\neq0$, it is straightforward to obtain the winding number for the effective
system described by Eq. (\ref{E19}), i.e.,%
\begin{equation}
C=\frac{1}{2}\text{sign}\left(  M\left(  z\right)  \right)  .
\end{equation}
This non-integral winding number appears since the deviations from this
two-band approximation model at large momenta are not included in the above
calculation of the winding number. So it can not be directly related to the
topological features of the system; however, the change in the winding numbers
is independent of the large-momentum contribution \cite{Qi}. Let us discuss
the change of the topological properties of superfluid system when we adjust
the oscillating amplitude $K$ of optical lattice. It is obvious that the
initial non-driven system is in a trivial state which corresponds to $z=0$ and
$M\left(  z\right)  <0$. Now, we apply the driven field to the system and make
$M\left(  z\right)  <0$ to $M\left(  z\right)  >0$, the change in Chern number
is:%
\begin{equation}
\Delta C=\frac{1}{2}\left[  \text{sign}\left(  M\left(  z\right)
_{>0}\right)  -\text{sign}\left(  M\left(  z\right)  _{<0}\right)  \right]
=+1.
\end{equation}
So, we get a topological superfluid state with $C=+1$ for $M\left(  z\right)
>0$ .

\begin{figure}[ptb]
\begin{center}
\includegraphics[width=1.0\linewidth]{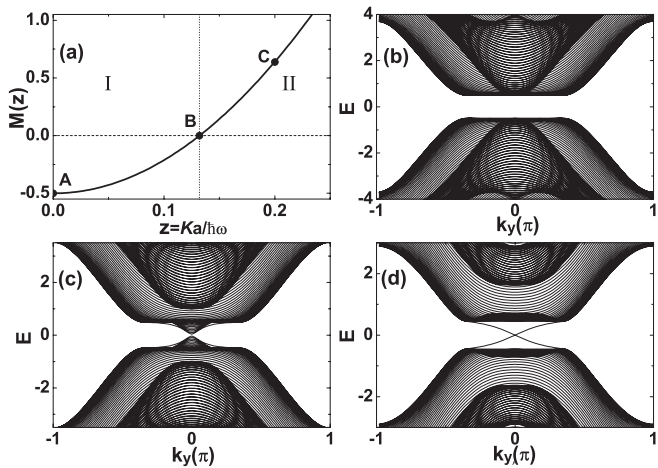}
\end{center}
\caption{ The phase diagram and the band structures of the system. (a) The mass $M\left(
z\right)  $ as a function of $z$. The region $I$ with $\left(  z\right)  < 0$
is a trivial superfluid, while the region $II$ with $M\left(  z\right)  > 0$
is a topological superfluid. The band structures of the effective Hamiltonian
(\ref{E10}) in a striped geometry with 60 sites in $x$ direction are shown in
(b), (c) and (d) corresponding to the points A ($z=0$), B ($z=0.132$), and C
($z=0.2$) in (a), respectively. Other parameters $t=1,\lambda=0.6,\psi
_{s}=-0.5,\hbar\omega=0.05,$ and $\mu=-4$.}
\end{figure}

It is notable that gapless chiral edge states are usually the hallmark of a
topological system. Therefore, to further prove the above argument, we show
the phase diagram and the band structures of the effective Hamiltonian
(\ref{E10}) in a striped geometry in Fig. 1. In Fig. 1(a), we plot the mass
$M\left(  z\right)  $ as a function of $z$, which can be adjusted by changing
the modulation strength $K$. The point \textbf{B} denotes $M\left(  z\right)
=0$ with $z=0.1320$, where the bands $E_{2}$ and $E_{3}$ contact each other at
$\Gamma$ point. The band structures in a striped geometry with $60$ sites in
$x$ direction are shown in (b), (c) and (d) corresponding to the points A
($z=0$), B ($z=0.132$), and C ($z=0.2$) in Fig.(a), respectively. In the
numerical calculation, we take the typical parameters $t=1,\lambda
=0.6,\psi_{s}=-0.5,\hbar\omega=0.05,$ and $\mu=-4$. It is clear that there is
no edge state in the region $I$ where the mass $M\left(  z\right)  <0$ with
$0\leq z<0.1320$, so it is a trivial superfluid. In contrast, there are a pair
of edge states in the region $II$ where the mass $M\left(  z\right)  >0$ with
$z>0.1320$, so it is a topological superfluid. There is a phase transition
from a trivial superfluid to a topological superfluid occurs at the point
$\mathbf{B}$ where the gap is closed.

Generally, there exists Majorana Fermionic excitation bounded with the vortex
structure in the nontrivial topological superfluid phase. Hence, we can obtain
the 2-D Foquet Majorana fermions \cite{Jiang} if there have the vortex
structures in our system. The vortex structure can be produced from two
different routes, one of which can be realized through the phase twist of the
SO-produced lasers: $\lambda\rightarrow\lambda e^{im\theta}$ with $m$ the
vorticity \cite{Sato}. Another route is that the vortex structure can come
from initial rotation of the atomic cloud \cite{Zwierlein}. Then the vortex
structure is coupled with the superfluid order parameter: $\psi_{s}%
\rightarrow\psi_{s}e^{im\theta}$, which is similar with the case in the
topological superconductor \cite{Sau}. Both cases give the similar Majorana
fermion obviously confirmed from the Eq. (\ref{E16}) and Eq. (\ref{E17})
connected by the unitary transformation $D$. The zero mode solutions of the
Majorana fermion can be obtained from the Bogoliubov-de Gennes (BdG) equation,
which has the similar form compared with that in Ref. \cite{Sato}. Moreover,
such Majorana fermion excitations can be detected by the standard Raman
spectroscopy \cite{Tewari,Zhu4}.

\section{CONCLUSION}

In summary, we have discussed the topological superfluid phase transition in a
periodically driven square optical lattice. By using Floquet's theorem, we
find that a Floquet topological superfluid will be created when the
two-dimensional square optical lattice potentials are periodically driven.
This topological phase is interesting in hosting a Majorana fermion excitation
which can be detected by Raman spectroscopy in cold atom system. Therefore we
propose a novel scenario to create Majorana fermions which may play a key role
in topological quantum computation.

\begin{acknowledgments}
G. Liu was supported by NSF of China (No.11147171), S. L. Zhu was supported in
part by the NBRPC (No.2011CBA00302), the SKPBRC (No.2011CB922104), and NSF of
China (No.11125417). This work was also supported by the NKBRSFC under grants
Nos. 2011CB921502, 2012CB821305, 2009CB930701, 2010CB922904, NSFC under grants
Nos. 10934010, 60978019, and NSFC-RGC under grants Nos. 11061160490 and 1386-N-HKU748/10.
\end{acknowledgments}


\begin{thebibliography}{99}                                                                                               %


\bibitem {Lewe}M. Lewenstein, A. Sanpera, V. Ahufinger, B. Damski, A. Sen(De),
U. Sen, Adv. Phys. \textbf{56}, 243 (2007).

\bibitem {Ruse}J. Ruseckas, G. Juzeli\={u}nas, P. \"{O}hberg, and M.
Fleischhauer, Phys. Rev. Lett. \textbf{95}, 010404 (2005).

\bibitem {Oste}K. Osterloh, M. Baig, L. Santos, P. Zoller, and M. Lewenstein,
Phys. Rev. Lett. \textbf{95}, 010403 (2005).

\bibitem {Juze}G. Juzeli\={u}nas, J. Ruseckas, P. \"{O}hberg, and M.
Fleischhauer, Phys. Rev. A \textbf{73}, 025602 (2006).

\bibitem {Zhu2}S. L. Zhu, H. Fu, C. J. Wu, S. C. Zhang, and L. M. Duan, Phys.
Rev. Lett. \textbf{97}, 240401 (2006); S. L. Zhu, D. W. Zhang, and Z. D. Wang,
Phys. Rev. Lett. \textbf{102}, 210403 (2009).

\bibitem {Lin1}Y. J. Lin, R. L. Compton, A. R. Perry, W. D. Phillips, J.V.
Porto, and I. B. Spielman, Phys. Rev. Lett. \textbf{102}, 130401 (2009).

\bibitem {Lin2}Y. J. Lin, R. L. Compton, K. Jim\'{e}nez-Garc\'{\i}a, J. V.
Porto, and I. B. Spielman, Nature \textbf{462}, 628 (2009).

\bibitem {Stanescu1}T. D. Stanescu, V. Galitski, J. Y. Vaishnav, C. W. Clark,
and S. Das Sarma, Phys. Rev. A \textbf{79}, 053639 (2009).

\bibitem {Bermudez}A. Bermudez, N. Goldman, A. Kubasiak, M. Lewenstein and M.
A. Martin-Delgado, New J. Phys. \textbf{12}, 033041 (2010).

\bibitem {Satija}I. I. Satija, D. C. Dakin, J. Y. Vaishnav, and C. W. Clark,
Phys. Rev. A \textbf{77}, 043410 (2008).

\bibitem {LXJ2}X. J. Liu, X. Liu, C. Wu, and J. Sinova, Phys. Rev. A
\textbf{81}, 033622 (2010).

\bibitem {Stan}T. D. Stanescu, V. Galitski, and S. D. Sarma, arXiv:0912.3559v1.

\bibitem {Mura}S. Murakami, Phys. Rev. Lett. \textbf{97}, 236805 (2006).

\bibitem {Koni}M. K\"{o}nig, S. Wiedmann, C. Bre\"{u}ne, A. Roth, H. Buhmann,
L. W. Molenkamp, X. L. Qi, S. C. Zhang, Science \textbf{318}, 766 (2007).

\bibitem {Hsieh3}D. Hsieh, D. Qian, L. Wray, Y. Xia, Y. S. Hor, R. J. Cava and
M. Z. Hasan, Nature \textbf{452}, 970 (2008).

\bibitem {Zhang}H. Zhang, C. X. Liu, X. L. Qi, X. Dai, Z. Fang, and S. C.
Zhang, Nature physics \textbf{5}, 438 (2009); X. L. Qi, R. Li, J. Zang, S. C.
Zhang, Science \textbf{323}, 1184 (2009).

\bibitem {Gold2}N. Goldman, I. Satija, P. Nikolic, A. Bermudez, M. A.
Martin-Delgado, M. Lewenstein, and I. B. Spielman, arXiv: 1002.0219v2; A.
Bermudez, L. Mazza, M. Rizzi, N. Goldman, M. Lewenstein, and M. A.
Martin-Delgado, arXiv: 1004.5101v1.

\bibitem {Zhu3}L. B. Shao, S. L. Zhu, L. Sheng, D.Y. Xing, and Z. D. Wang,
Phys. Rev. Lett. \textbf{101}, 246810 (2008).

\bibitem {Liu}G. Liu, S. L. Zhu, S. Jiang, F. Sun, and W. M. Liu, Phys. Rev. A
\textbf{82}, 053605 (2010).

\bibitem {Fu}L. Fu and C. L. Kane, Phys. Rev. B \textbf{74}, 195312 (2006).

\bibitem {Inoue1}J. I. Inoue, Phys. Rev. B \textbf{81}, 125412 (2010).

\bibitem {Thouless}D. J. Thouless, Phys. Rev. B \textbf{27}, 6083 (1983).

\bibitem {Niu}Q. Niu and D. J. Thouless, J. Phys. A \textbf{17}, 2453 (1984);
S. L. Zhu and Z. D. Wang, Phys. Rev. B \textbf{65}, 155313 (2002)

\bibitem {Ecka}A. Eckardt, C. Weiss, and M. Holthaus, Phys. Rev. Lett.
\textbf{95}, 260404 (2005).

\bibitem {Jiang}L. Jiang, T. Kitagawa, J. Alicea, A. R. Akhmerov, D. Pekker,
G. Refael, J. I. Cirac, E. Demler, M. D. Lukin, and P. Zoller, Phys. Rev.
Lett. \textbf{106}, 220402 (2011).

\bibitem {Kita}T. Kitagawa, E. Berg, M. Rudner, and E. Demler, Phys. Rev. B
\textbf{82}, 235114 (2010).

\bibitem {Inoue2}J. I. Inoue and A. Tanaka, Phys. Rev. Lett. \textbf{105},
017401 (2010).

\bibitem {Lind}N. H. Lindner, G. Refael and V. Galitski, Nature Physics
\textbf{7}, 490 (2011).

\bibitem {Kita2}T. Kitagawa, L. Fu, E. Demler, T. Oka and A. Brataas, arXiv: 1104.4636v1.

\bibitem {Sato}M. Sato, Y. Takahashi, and S. Fujimoto, Phys. Rev. Lett.
\textbf{103}, 020401 (2009).

\bibitem {ZhangPRL}C. Zhang, S. Tewari, R. M. Lutchyn, and S. Das Sarma, Phys.
Rev. Lett. \textbf{101}, 160401 (2008)

\bibitem {ZhangPRA}C. Zhang, Phys. Rev. A \textbf{82}, 021607(R) (2010).

\bibitem {Zhu4}S. L. Zhu, L. B. Shao, Z. D. Wang, and L. M. Duan, Phys. Rev.
Lett. \textbf{106}, 100404 (2011).

\bibitem {Shirl}J. H. Shirley, Phys. Rev. \textbf{138}, B979 (1965).

\bibitem {Grif}M. Grifoni, P. H\"{a}nggi, Physics Reports \textbf{304}, 229 (1998).

\bibitem {Qi}X. L. Qi and S. C. Zhang, Rev. Mod. Phys. \textbf{83}, 1057 (2011).

\bibitem {Zwierlein}M. W. Zwierlein, J. R. Abo-Shaeer, A. Schirotzek, C. H.
Schunck, W. Ketterle, Nature (London) \textbf{435}, 1047 (2005).

\bibitem {Sau}J. D. Sau, R. M. Lutchyn, S. Tewari, and S. Das Sarma, Phys.
Rev. Lett. 104, 040502 (2010).

\bibitem {Tewari}S. Tewari, S. Das Sarma, C. Nayak, C. Zhang, and P. Zoller,
Phys. Rev. Lett. \textbf{98}, 010506 (2007).
\end{thebibliography}
\end{document}